# Hyperbolic phonon polaritons and wave vector direction dependent dielectric tensors in anisotropic crystals

*Yue Fang*[a], *Huanjun Chen*[b], *Zhibing Li*[c], *and Weiliang Wang* [a]*

[a]*School of Physics, Guangdong Province Key Laboratory of Display Material and Technology, Center for Neutron Science and Technology, Sun Yat-sen University, Guangzhou 510275, China*

[b]*School of Electronics and Information Technology, State Key Laboratory of Optoelectronic Materials and Technologies, Guangdong Province Key Laboratory of Display Material and Technology, Sun Yat-sen University, Guangzhou 510275, China ;*

[c]*School of Science, State Key Laboratory of Optoelectronic Materials and Technologies, Guangdong Province Key Laboratory of Display Material and Technology, Shenzhen Campus of Sun Yat-sen University Shenzhen 518107, China*

[*]*Email: wangwl2@mail.sysu.edu.cn*

**Abstract:** Hyperbolic phonon polariton is important in precisely controlling photons at the nanoscale. It was common practice to calculate the dielectric function of the phonon polariton system with the Drude-Lorenz model. We considered the impact of LO-TO splitting while applying the Drude-Lorenz model. Then the dielectric functions become wave vector direction dependent besides electric polarization direction dependent. Our results show that considering LO-TO splitting can more accurately predict dielectric functions. Additionally, we discovered that, besides hexagonal BN, hexagonal AlN exhibits a wide hyperbolic frequency band range, while the other four Ⅲ-Ⅴ semiconductor materials(BP, AlP, GaN, GaP) display it scarcely. Furthermore, we found that the phonon frequency, lifetime, and the difference of infrared active transverse optical phonon frequencies with different wave vector directions are critical factors in determining the width of the hyperbolic frequency band range. We also found some dumbbell-shaped and butterfly-shaped isofrequency curves in h-AlN, h-GaP, and especially h-GaN. Our study provides a fresh perspective on understanding the dielectric properties of these materials and lays a theoretical foundation for further exploration and development of new hyperbolic phonon polariton materials.

## TOC GRAPHICS

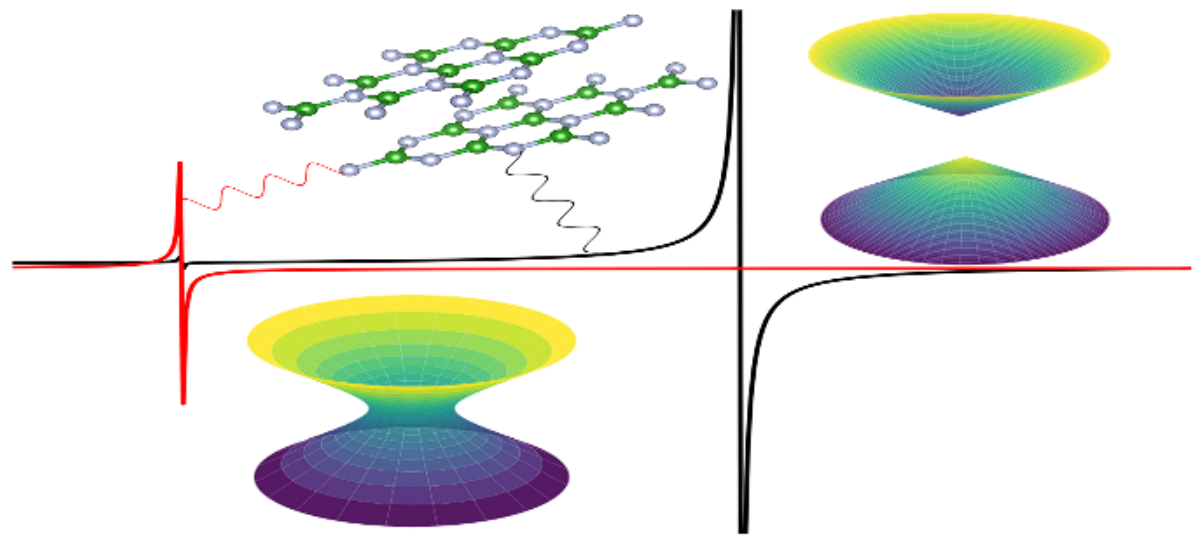

## 1. Introduction

Recently, the concept of photonic chips emerged as photons can carry more information and transport faster than electrons. This makes photonic chips ideal for constructing next-generation low-energy-consumption, high-density, and high-efficiency information devices. Manipulating photons at the nanoscale is challenging, as photons do not carry charge and have weak interactions with matter. However, phonon polaritons (PhPs) possess both optical phonon and photon properties, making them crucial in the precise control of photons at the nanoscale. [1]

Some polar crystals exhibit fascinating properties attributed to the existence of phonon polaritons. When subjected to a particular frequency range, the product of the dielectric functions in distinct directions becomes negative, leading to a remarkable level of anisotropy.[2] This occurrence resembles the characteristics of hyperbolic materials. Hyperbolic materials are usually uniaxial or biaxial, as an anisotropic material. The relationship between wave vector and frequency of the extraordinary wave in uniaxial crystals is [1–3]

$$\frac{k_x^2 + k_y^2}{\varepsilon_z} + \frac{k_z^2}{\varepsilon_t} = \left(\frac{\omega}{c}\right)^2 \tag{1}$$

The dielectric function tensor comprises two key components: $\varepsilon_z$ and $\varepsilon_t$. The former is the dielectric function when the electric field polarization is parallel to the optic axis. Conversely, the latter is the dielectric function when the electric field polarization is perpendicular to the optic axis. When $\varepsilon_t$ and $\varepsilon_z$ have opposite signs, the isofrequency surface in the wave vector space is a hyperboloid, which features several unique properties compared to conventional materials whose isofrequency surface is an ellipsoid. Hyperbolic materials can exhibit dielectric and metallic properties along different crystal axes and can be classified into two types based on the signs of their dielectric functions. Materials with $\varepsilon_t > 0$ and $\varepsilon_z < 0$ primarily possess dielectric properties, also known as dielectric or type I hyperbolic materials. Conversely, materials with $\varepsilon_t < 0$ and $\varepsilon_z > 0$ mainly exhibit metallic properties, referred to as metallic or type II hyperbolic materials. [1]

The distinctive hyperbolic isofrequency surfaces have numerous new characteristics and applications, offering a tremendous opportunity to develop new technologies in the mid-infrared (mid-IR) for applications including enhanced and modulated sensing through vibrational spectroscopy, [4,5] enhanced radiative heat transfer, [6–8] enhanced spontaneous emission,[9,10] negative refraction [11–13] and other optical applications. [14,15] While artificial hyperbolic metamaterials have been demonstrated, they suffer from high plasmonic losses and require complicated nanofabrication, which in turn induces size-dependent limitations on optical confinement. The low-loss, natural, hyperbolic PhP material is an attractive alternative [3,16] , and there are various modulation methods.[17,18] They simplify the synthesis of microstructures and support higher electromagnetic confinement and photon state densities [3]. Extensive research has been conducted in natural hyperbolic PhP materials, including hexagonal boron nitride (h-BN), [19] α-$MoO_3$, [20] calcite, [21] and $Ca_2N$. [11] Our study aimed to search for more natural hyperbolic PhP materials with wide hyperbolic bands. We initiated our investigation by examining the h-BN structure. Then, we analyzed the dielectric functions of hexagonal III-V semiconductors (BN, BP, AlN, AlP, GaN, GaP) that share similar structures (Figure 1).

There are two types of Drude-Lorenz model widely used in calculating the dielectric functions of materials containing PhPs. One is called LOTO formalism or four-parameter semi-quantum model [2,22–24] :

$$\varepsilon_a(\omega) = \varepsilon_{a,\infty} \prod_j \frac{\left(\omega_{LO,j}^a\right)^2 - \omega^2 + i\gamma_{LO,j}^a\omega}{\left(\omega_{TO,j}^a\right)^2 - \omega^2 + i\gamma_{TO,j}^a\omega} \tag{2}$$

where $a = x, y, z$, $\omega_j$ and $\gamma_j$ denote the frequency and the inversion of lifetime of the $j$-th IR-active phono mode, respectively. $\varepsilon_{a,\infty}$ is the high-frequency dielectric constant. LO and TO denote the longitudinal and transverse IR phonon modes, respectively. $j$ goes over all the IR-active modes. However, this expression does not contain the different dipole moments of each IR phonon mode and confuses the wave vector direction with the electric field polarization direction. The other type of expression of the Drude-Lorenz model which contains the oscillator strength ($S_{j,\alpha\beta}$) is better [25–28] :

$$\varepsilon_{\alpha\beta}(\omega) = \varepsilon_{\alpha\beta}^{\infty} + \frac{4\pi}{\Omega} \sum_j \frac{S_{j,\alpha\beta}}{\omega_j^2 - \omega^2 + i\gamma_j\omega} \tag{3}$$

where $\varepsilon_{\alpha\beta}^{\infty}$ is the electronic dielectric permitttivity tensor element, α and β are the indices of directions, Ω is the primitive cell volume. The phonon frequencies needed in the Drude-Lorenz model can be calculated through first-principles calculations [26,29,30] or fitted from experimental results. [12,30,31] After determining the dielectric functions, researchers can assess the imaginary part of p-polarized (where the electric field polarization direction is within the plane of incidence) infrared light's reflectance [32–35] to gain insight into the dispersion relations of phonon polaritons.

This work also employs the second type expression of the Drude-Lorenz model to compute the dielectric function. What sets our methodology apart is the inclusion of the effects of LO-TO splitting, [26,36,37] which is integral to achieving precise analysis. Without considering LO-TO splitting, the phonon dispersion at the $\Gamma$ point is continuous. However, the continuity is disrupted when LO-TO splitting is factored in. This results in phonons near the $\Gamma$ point displaying varying frequencies depending on their wave vector directions (Figure 2). Thus, the dielectric function depends not only on the direction of electric field polarization but also on the wave vector direction. Although the dielectric function in Equation (1) also appears to depend on the wave vector direction, it actually depends only on the electric field polarization. Because $\varepsilon_t$ and $\varepsilon_z$ do not depend on the wave vector direction, it differs distinctly from the wave vector direction dependence of dielectric function due to LO-TO splitting in which $\varepsilon_t$ and $\varepsilon_z$ depend on the wave vector direction.

## 2. Method

### 2.1 Dielectric function of a phonon (phonon polariton) system

The potential energy of the phonon system is a function of atomic positions:

$$V[\vec{r}(j_1 l_1), \cdots, \vec{r}(j_n l_N)], \tag{4}$$

where $\vec{r}(jl)$ is the displacement of the *j*-th atom in the *l*-th unit cell, and *n* and *N* are the number of atoms in a unit cell and the number of unit cells, respectively. When an atom has a displacement in the direction of α, it will feel a force

$$F_\alpha(jl) = -\frac{\partial V}{\partial r_\alpha(jl)}. \tag{5}$$

This first order derivative is zero at the equilibrium position. Therefore, it is necessary to take into account the second force constant

$$\Phi_{\alpha\beta}(jl, j'l') = \frac{\partial^2 V}{\partial r_\alpha(jl)\partial r_\beta(j'l')} = -\frac{\partial F_\beta(j'l')}{\partial r_\alpha(jl)} \tag{6}$$

Then, the force is

$$F_\beta(j'l') = m_{j'}\ddot{r}_\beta(j'l') = -\sum_{\alpha jl} \Phi_{\alpha\beta}(jl, j'l') * r_\alpha(jl) + eZ_{\gamma\beta}(j'l') * E_\gamma. \tag{7}$$

The second term on the right-hand side is the electric force, where *E* is the electric field, *e* is the electron charge, *Z*(*j'l'*) is the Born effective charge tensor of this atom

$$Z_{\alpha\beta} = \left.\frac{\partial^2 V}{\partial E_\alpha \partial r_\beta}\right|_{E=0, r=0} = \left.\frac{\partial F_\beta}{\partial E_\alpha}\right|_{u=0} = \left.\frac{\partial p_\alpha}{\partial r_\beta}\right|_{E=0} \tag{8}$$

i.e. the force on the atom is

$$F_\beta = \sum_\alpha Z_{\alpha\beta} E_\alpha, \tag{9}$$

The electric dipole moment induced by atom displacement is

$$p_\alpha = \sum_\beta Z_{\alpha\beta} r_\beta \tag{10}$$

According to the Bloch theorem, the solution of **Equation (7)** has the form

$$r_\alpha(jl) = \frac{e_\alpha(j, \vec{q}) e^{-i\omega(\vec{q})t} e^{i\vec{q}\cdot\vec{r}(jl)}}{\sqrt{m_j}}. \tag{11}$$

Consider an oscillating electric field

$$E_\gamma = E_{\gamma 0} e^{-i\omega_L t} \tag{12}$$

Substitute **Equation (11)** into Equation (7), we get

$$\ddot{e}_\alpha(j,\vec{q},t) = \sum_{\beta j'} \mathrm{D}_{\alpha\beta}(jj',q) e_\beta(j',\vec{q},t) - \frac{e^{-i\vec{q}\cdot\vec{r}(j0)}}{\sqrt{m_j}} eZ_{\gamma\alpha}(j0) * E_{\gamma 0} e^{-i\omega_L t} \quad (13)$$

where

$$D_{\alpha\beta}(jj',q) = \frac{1}{\sqrt{m_j m_{j'}}} \sum_{l'} \Phi_{\alpha\beta}(j0,j'l') \exp\{i\vec{q}\cdot[\vec{r}(j'l') - \vec{r}(j0)]\} \quad (14)$$

Assume that the $i$-th eigen vector of D is $Q_i d_i(\alpha j,\vec{q})$, where $d_i(\alpha j,\vec{q})$ is the normalized eigenvector, $Q_i$ is the amplitude, the corresponding eigen value is $[\omega_p(\vec{q})]^2$. The Equation (13) becomes

$$\ddot{Q}_i d_i(\alpha j,\vec{q}) = -[\omega_p(\vec{q})]^2 Q_i d_i(\alpha j,\vec{q}) + \frac{e^{-i\vec{q}\cdot\vec{r}(j0)}}{\sqrt{m_j}} eZ_{\sigma\alpha}(j0) * E_{\sigma 0} e^{-i\omega_L t} \quad (15)$$

If the eigenstate's lifetime is $1/\gamma$, then Equation (15) should be rewritten as

$$\ddot{Q}_i d_i(\alpha j,\vec{q},t) + \gamma Q_i d_i(\alpha j,\vec{q}) + [\omega_p(\vec{q})]^2 Q_i d_i(\alpha j,\vec{q}) = \frac{e^{-i\vec{q}\cdot\vec{r}(j0)}}{\sqrt{m_j}} eZ_{\sigma\alpha}(j0) * E_{\sigma 0} e^{-i\omega_L t} \quad (16)$$

The $Q_i$ satisfies

$$\ddot{Q}_i + \gamma Q_i + [\omega_p(\vec{q})]^2 Q_i = \sum_{\alpha j} \frac{e^{-i\vec{q}\cdot\vec{r}(j0)}}{\sqrt{m_j}} eZ_{\sigma\alpha}(j0) d_i^*(\alpha j,\vec{q},t) * E_{\sigma 0} e^{-i\omega_L t} \quad (17)$$

Its solution is

$$Q_i = \frac{1}{[\omega_p(\vec{q})]^2 - \omega_L^2 - i\gamma\omega_L} \sum_{\alpha j} \frac{e^{-i\vec{q}\cdot\vec{r}(j0)}}{\sqrt{m_j}} eZ_{\sigma\alpha}(j0) d_i^*(\alpha j,\vec{q},t) * E_{\sigma 0} e^{-i\omega_L t} \quad (18)$$

The electric polarization induced by these atom displacements is

$$\begin{aligned} P_\beta &= \frac{e}{\Omega} \sum_{\mu j'} Z_{\beta\mu} r_\mu(j'0,t) = \frac{e}{\Omega} \sum_{\mu j' k} Z_{\beta\mu} \frac{Q_k d_k(\mu j',\vec{q},t) e^{i\vec{q}\cdot\vec{r}(j'0)}}{\sqrt{m_{j'}}} \\ &= \frac{e}{\Omega} \frac{1}{[\omega_p(\vec{q})]^2 - \omega_L^2 - i\gamma\omega_L} \sum_{\mu j' k} \frac{Z_{\beta\mu}(j')}{\sqrt{m_{j'}}} d_k(\mu j',\vec{q},t) \\ &\quad * \sum_{\alpha j} \frac{e^{i\vec{q}\cdot\vec{r}(j'0)} e^{-i\vec{q}\cdot\vec{r}(j0)}}{\sqrt{m_j}} eZ_{\sigma\alpha}(j) d_k^*(\alpha j,\vec{q},t) * E_{\sigma 0} e^{-i\omega_L t} \end{aligned} \quad (19)$$

where $\Omega$ is the volumn of a unit cell. When the wave length is much longer that the size of the unit cell, $q \sim 0$, then the polarizability.

$$\chi_{\beta\sigma} = \frac{P_\beta}{\varepsilon_0 E_\sigma} = \frac{e^2}{\varepsilon_0 \Omega} \sum_{\mu j' k} \frac{1}{\left[\omega_{pk}(\vec{q})\right]^2 - \omega_L^2 - i\gamma\omega_L} \frac{Z_{\beta\mu}(j')}{\sqrt{m_{j'}}} d_k(\mu j', \vec{q}, t) * \sum_{\alpha j} \frac{Z_{\sigma\alpha}(j)}{\sqrt{m_j}} d_k^*(\alpha j, \vec{q}, t) \quad (20)$$

In the long-wavelength limit ($\vec{q} \approx 0$), the polarizability is Equation (20) (similar to Ref. [25–28]), where $\omega_{pk}(\vec{q})$ is the frequency of the $k$-th phonon near the $\Gamma$ point along a certain direction. The optical phonon is coupled with the electromagnetic wave, therefore, $\omega_{pk}$ is actually the frequency of the PhP. The dielectric function tensor $\overleftrightarrow{\varepsilon} = \overleftrightarrow{\chi}\varepsilon_0 + \overleftrightarrow{\varepsilon}_\infty$ is then dependent on the phonon frequency, phonon lifetime, and dipole moment of the phonon mode, which can be obtained through first principle calculation.

**2.2 Refraction law in anisotropic material whose dielectric function depends on wave vector direction.**

In the material

$$P" = P_0" \exp[i(\vec{q} \cdot \vec{r} - \omega t)] \quad (21)$$

$$E" = E_0" \exp[i(\vec{q} \cdot \vec{r} - \omega t)] \quad (22)$$

$$H" = H_0" \exp[i(\vec{q} \cdot \vec{r} - \omega t)] \quad (23)$$

Then the Maxwell' equations become

$$\vec{q} \times \vec{E}_0" = \mu_0 \omega \vec{H}_0" \quad (24)$$

$$\vec{q} \times \vec{H}_0" = -\omega \vec{D}_0" \quad (25)$$

$$\vec{q} \cdot \vec{D}_0" = 0 \quad (26)$$

$$\vec{q} \cdot \vec{H}_0" = 0 \quad (27)$$

Let the $z$-axis perpendicular to the interface between the material and the air, and $x$-$z$ plane be the incident plane, and the angle between $\vec{q}$ and the $z$-axis is $\theta"$. As $\vec{D}_0"$ is perpendicular to $\vec{q}$, $D_z" = \tan(\theta")D_x"$.

In principal axes system, the dielectric function tensor is diagonalized, then the electric field $\vec{E}_0"=\varepsilon_x^{-1}(\theta")D_x"\vec{e}_x + \varepsilon_z^{-1}(\theta")D_z"\vec{e}_z$, whose component perpendicular to $\vec{q}$ is $E_{\perp q}"=\left|\vec{e}_q \times \vec{E}_0"\right|$, where $\vec{e}_q$ is the unit vector in $\vec{q}$ direction. Equation (24) leads to

$$qE_{\perp q}" = \mu_0 \omega H_0" \quad (28)$$

Equation (25) leads to

$$qH_0" = \omega D_0" \quad (29)$$

After eliminating $H_0^{''}$ in Equation (28-29), we get

$$q = \sqrt{\frac{\mu_0 \omega^2 D_0^{''}}{E_{\perp q}^{''}}} \tag{30}$$

Then, we can get the relation between the incident angle $\theta$ and the refraction angle $\theta^{''}$ by substituting Equation (30) into the conventional refraction law

$$k\sin\theta = q\sin\theta^{''} \tag{31}$$

where $k = \frac{\omega}{c}$ is the wave vector in the air. The $c$ is the speed of light in the air. Equation (30) also gives us $q(\theta'')$ which is the isofrequency curve.

**2.3 Density functional theory calculation**

We calculated the electronic structure and the force constant matrix with density functional theory implemented in the Vienna Ab initio Simulation Package (VASP).[38–40] Then, the phonon dispersion and the eigenmodes are obtained with the software package Phonopy.[41,42] The phonon lifetime is obtained with the phono3py package.[41,42] Density functional theory calculations utilized the generalized gradient approximation (GGA) of Perdew-Burke-Ernzerhof (PBE)[43] for the exchange-correlation functional, with a plane wave cutoff energy set at 520 eV. Our energy convergence precision was set at $10^{-10}$ eV/cell, while the force convergence accuracy applied to each atom was established at $10^{-8}$ eV/Å. We computed the Born effective charge tensors using the density functional theory-linear response scheme, coupled with the iterative Green's function approach for density-functional perturbation theory.[26] The $k$-point mesh and supercell size are provided in Table 1.

Table.1: The $k$-point mesh and supercell size in DFT and phonon calculation

| | *k*-point mesh | | | | Supercell size | |
|---|---|---|---|---|---|---|
| | Structure optimization | Born effective charge calculation | Force constant calculation | Phonon lifetime calculation | Force constant calculation | Phonon lifetime calculation |
| h-BN | 12×12×3 | 12×12×5 | 2×2×2 | 9×9×2 | 5×5×2 | 2×2×2 |
| h-BP | 12×12×5 | 12×12×5 | 3×3×2 | 2×2×2 | 3×3×3 | 2×2×2 |
| h-AlN | 12×12×12 | 12×12×12 | 3×3×2 | 4×4×4 | 3×3×3 | 2×2×2 |
| h-AlP | 9×9×9 | 12×12×5 | 2×2×2 | 2×2×2 | 4×4×5 | 2×2×2 |
| h-GaN | 7×7×4 | 12×12×5 | 2×2×2 | 5×5×2 | 4×4×2 | 2×2×2 |
| h-GaP | 6×6×3 | 9×9×4 | 2×2×2 | 4×4×2 | 4×4×2 | 2×2×2 |

Table.2: Lattice constants of optimized crystal structures (unit: Angstrom)

| Crystal | $a = b$ | $c$ |
|---|---|---|
| h-BN | 2.51 | 7.71 |
| h-BP | 3.20 | 5.30 |
| h-AlN | 3.31 | 4.18 |
| h-AlP | 3.89 | 6.38 |
| h-GaN | 3.20 | 5.30 |
| h-GaP | 3.83 | 6.31 |

## 3. Results and Discussion

In this paper, we investigated bulk h-BN and five other similar materials: h-BP, h-AlN, h-AlP, h-GaN, and h-GaP, all of which belong to the hexagonal crystal system. Figure 1 shows their optimized structures, and Table 2 details the lattice constants obtained after optimization. The structure of h-BP, h-AlP, h-GaN, and h-GaP are wurtzite structures. We calculated the phonon dispersion for these six bulk materials, and Figure 2 shows that they exhibit no imaginary frequencies, indicating their dynamical stability. It is worth noting that the phonon dispersion of h-BN, h-AlN, and h-GaN are similar to those of the bulk materials described in Ref. [44,45] and the monolayer structures detailed in Ref. [46].

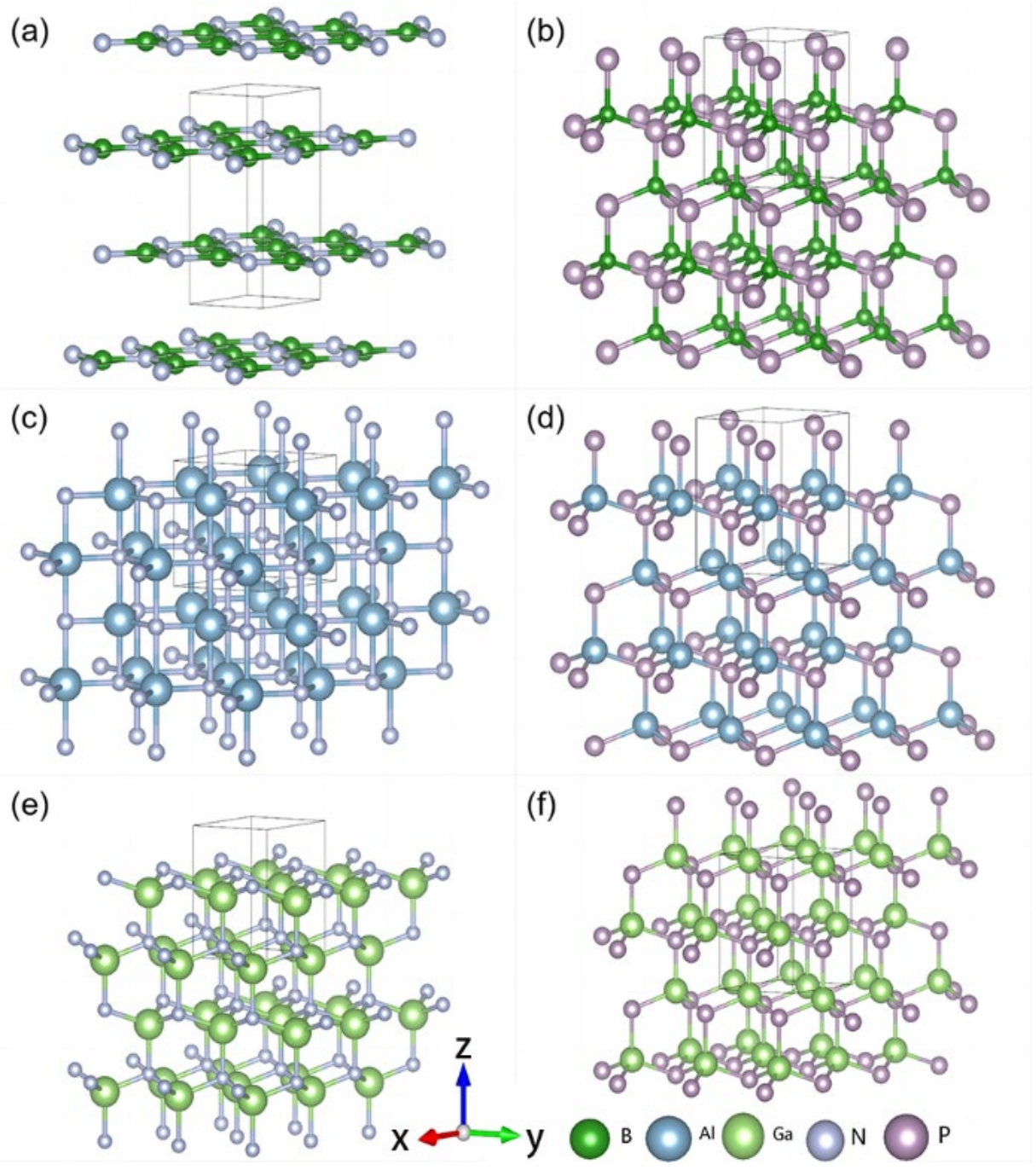

Fig. 1: Atomic structure of (a) h-BN, (b) h-BP, (c) h-AlN, (d) h-AlP, (e) h-GaN, and (f) h-GaP.

Based on group theory analysis, the irreducible representations at the Γ point for bulk h-BN and h-AlN are denoted as $2E_{2g} + 2B_{2g} + 2A_{2u} + 2E_{1u}$, where $E_{2g}$ and $E_{1u}$ represent doubly degenerate modes, and $B_{2g}$ and $A_{2u}$ represent non-degenerate modes. Among these, $2A_{2u} + 2E_{1u}$ exhibit IR activity. For bulk h-BP, h-AlP, h-GaN, and h-GaP, the irreducible representations at the $\Gamma$ point are given by $2A_1 + 2B_1 + 2E_2 + 2E_1$, where $E_2$ and $E_1$ are doubly degenerate modes, and $B_1$ and $A_1$ are non-degenerate modes. Here, $2A_1 + 2E_1$ are IR-active. Since only the IR-active phonons influence the dielectric function, we have labeled the IR-active phonons with tags in red in Figure 2

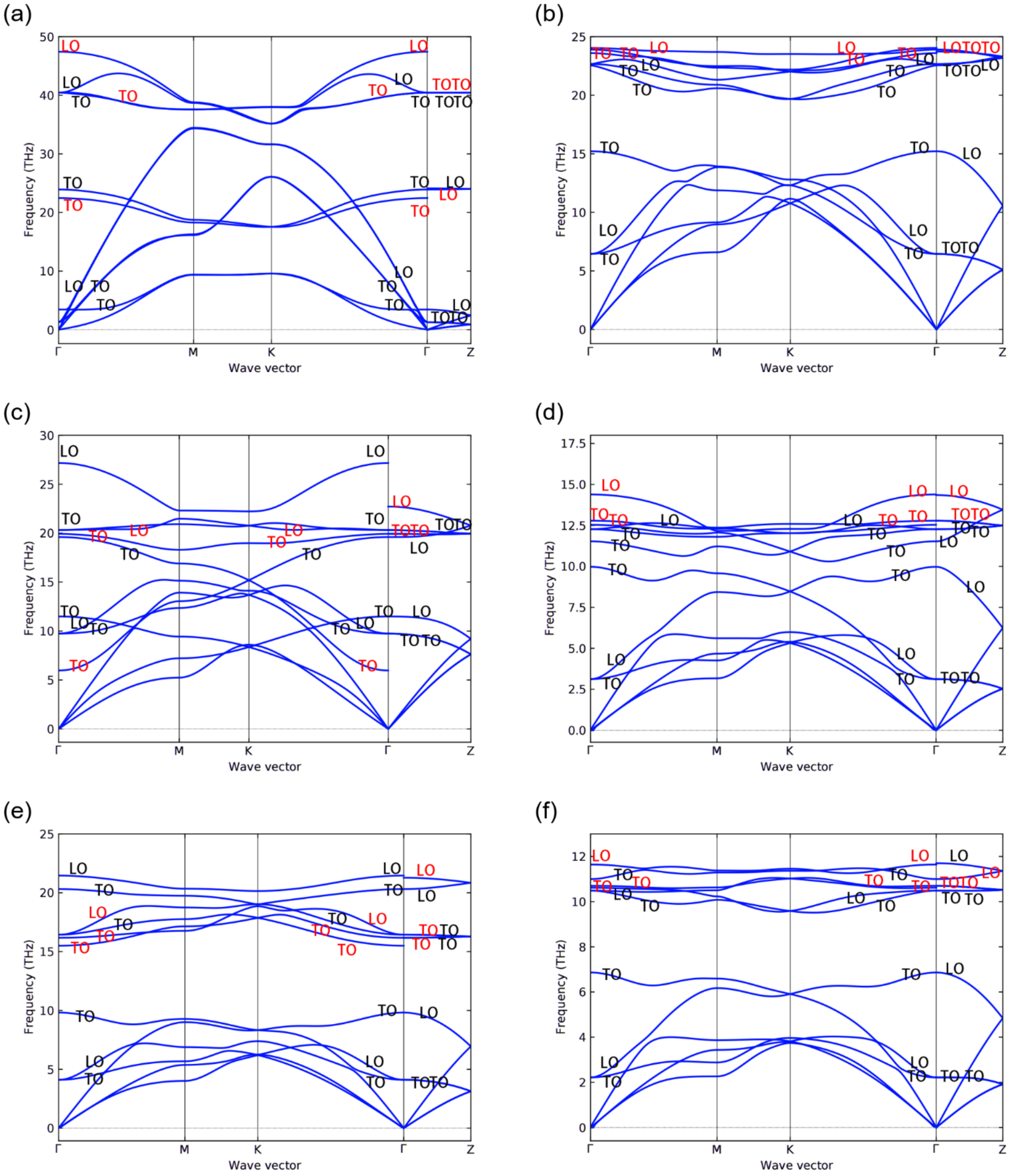

Fig. 2: Phonon dispersion of (a) h-BN, (b) h-BP, (c) h-AlN, (d) h-AlP, (e) h-GaN, and (f) h-GaP. Infrared active modes are labeled with red tags.

Firstly, we thoroughly analyzed the dielectric function of h-BN and compared it to previously published experimental data [3] (Figure 3a). Our theoretical results align with the experimental outcomes more than the theoretical results in Ref.[10].This is due to our consideration of the variation of phonon frequencies in different wave vector directions that result from LO-TO splitting. The wave vector direction we considered is the same as in the Ref. [3] experiment. The incident wave vector forms a 25-degree angle with the z-axis[22] and then the refraction wave vector direction is determined self-consistently. The refraction angle (the wave vector direction inside h-BN) obtained with calculation method section 2.2 is shown in Figure 4 .

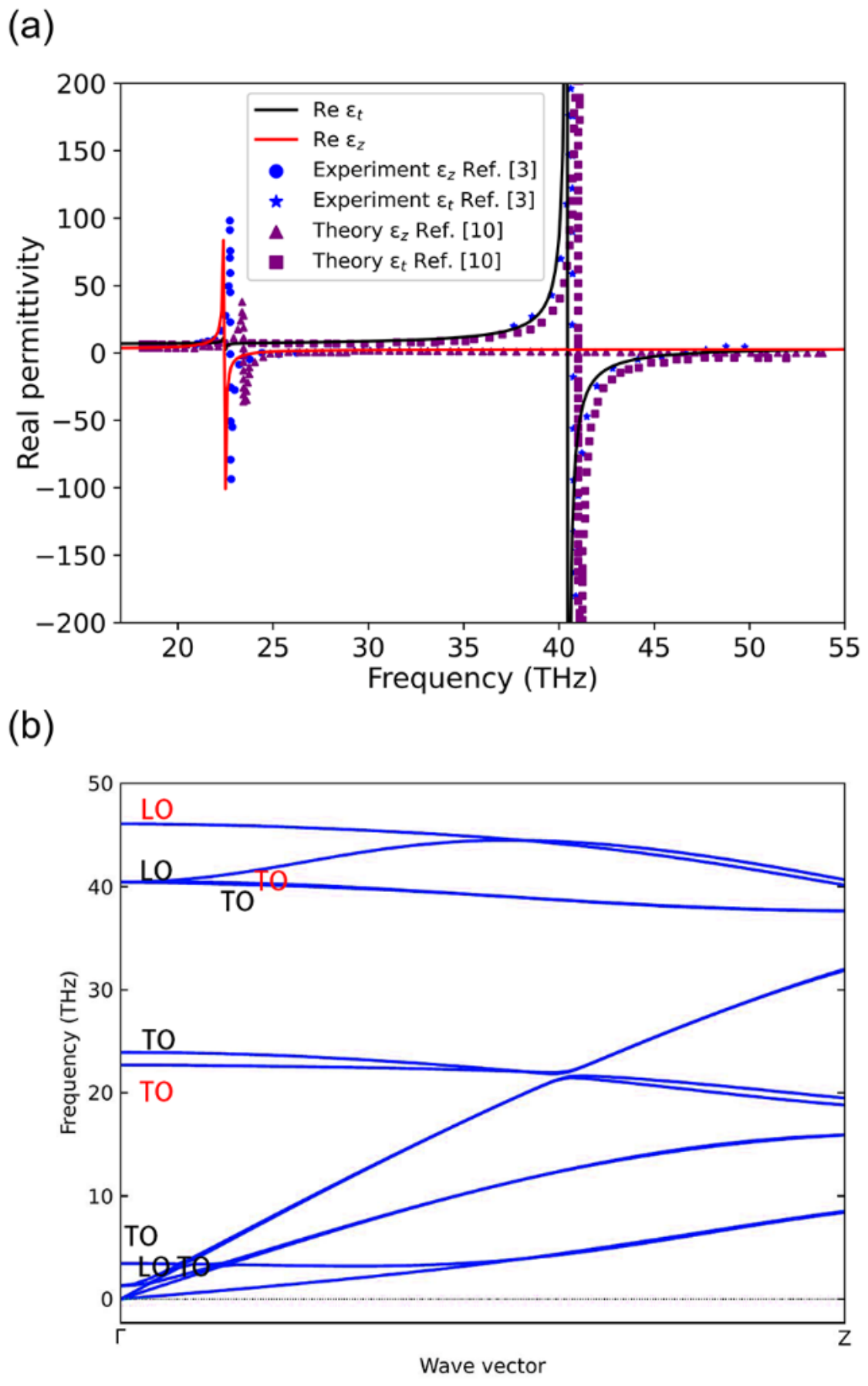

Fig. 3: (a) The real part of the dielectric function in h-BN when the incident wave vector is along a direction forming a 25° angle with the *z*-axis within the *x*-*z* plane (The refraction angle is shown in Figure 4). The black line refers to the dielectric function with polarization direction in the *x*-*y* plane; the red line refers to the dielectric function with polarization along the *z*-axis. The blue circles and stars are the experimental data from Ref.[3], and the purple triangles and squares are the theoretical results from Ref.[10] (b) Phonon dispersion of h-BN when the wave vector is in the *x*-*z* plane and forms an angle of 25° with the *z*-axis. Infrared active modes are labeled with red tags.

To roughly explain the results in Figure 3a, we plot the phonon dispersion with the wave vector direction forms a 25-degree angle with the *z*-axis (this angle should depend on the laser frequency) in Figure 3b. We can identify two IR-active transverse optical (TO) phonon frequencies at 23.0 THz and 40.4 THz. As a rough estimation, these two phonon frequencies correspond to a peak in the dielectric function at these frequencies (Figure 3a), consistent with the experimental data. Had we not considered the LO-TO splitting, the IR-active TO frequencies would have been 23.3 THz and 40.5 THz (see phonon dispersion in Fig. 5), leading to discrepancies between the peak positions of the dielectric function and the experimental data. If we take the wave vector direction along the *x*-axis or *z*-axis in h-BN, the dielectric function (Figure 6a) is also very different from the experimental results, so it is essential to consider the variation of phonon frequencies in different wave vector directions caused by LO-TO splitting. Note that our results in Figure 3a are not derived from Figure 3b, in which the wave vector direction forms a constant 25-degree angle with the *z*-axis. Instead, the wave vector direction inside the media (refraction wave vector direction) and the dielectric function tensor are determined self-consistently with a given laser frequency and incident angle (25 degrees). The refraction angle corresponding to the dielectric function in Figure 3a is shown in figure 4. The refraction angle is obtained with the calculation method section 2.2

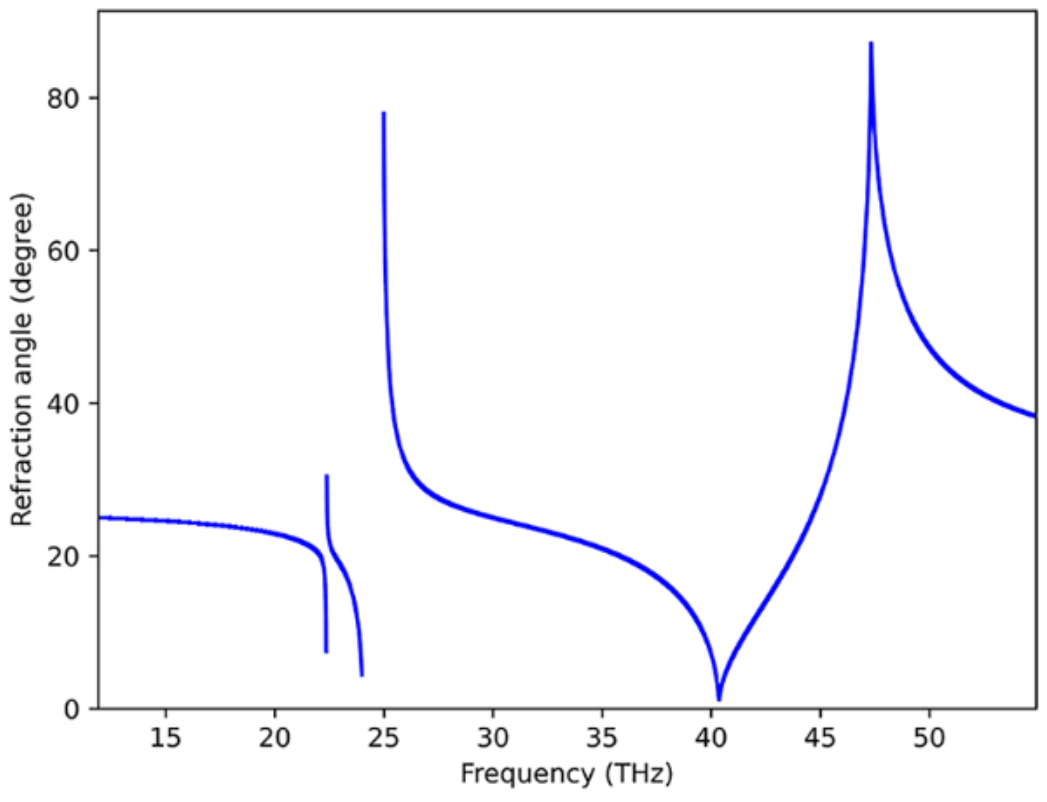

Fig. 4: Refraction angle in h-BN with incident angle 25 degrees.

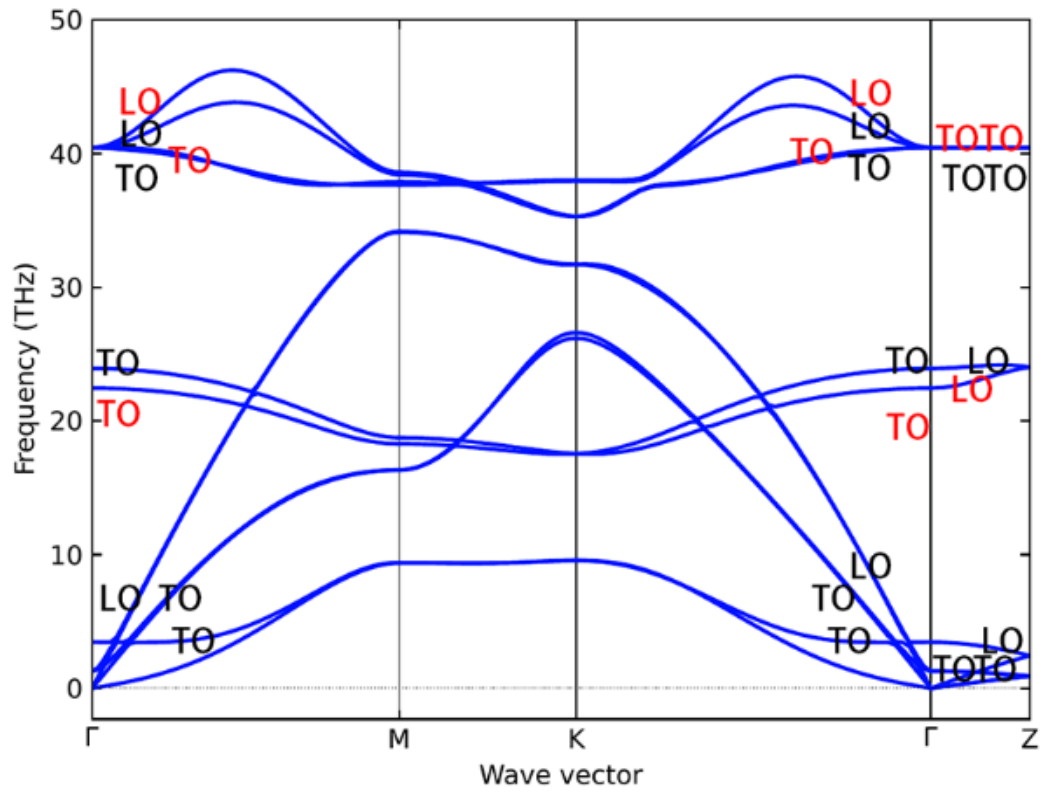

Fig. 5: Phonon dispersion of h-BN without taking LO-TO splitting into account.

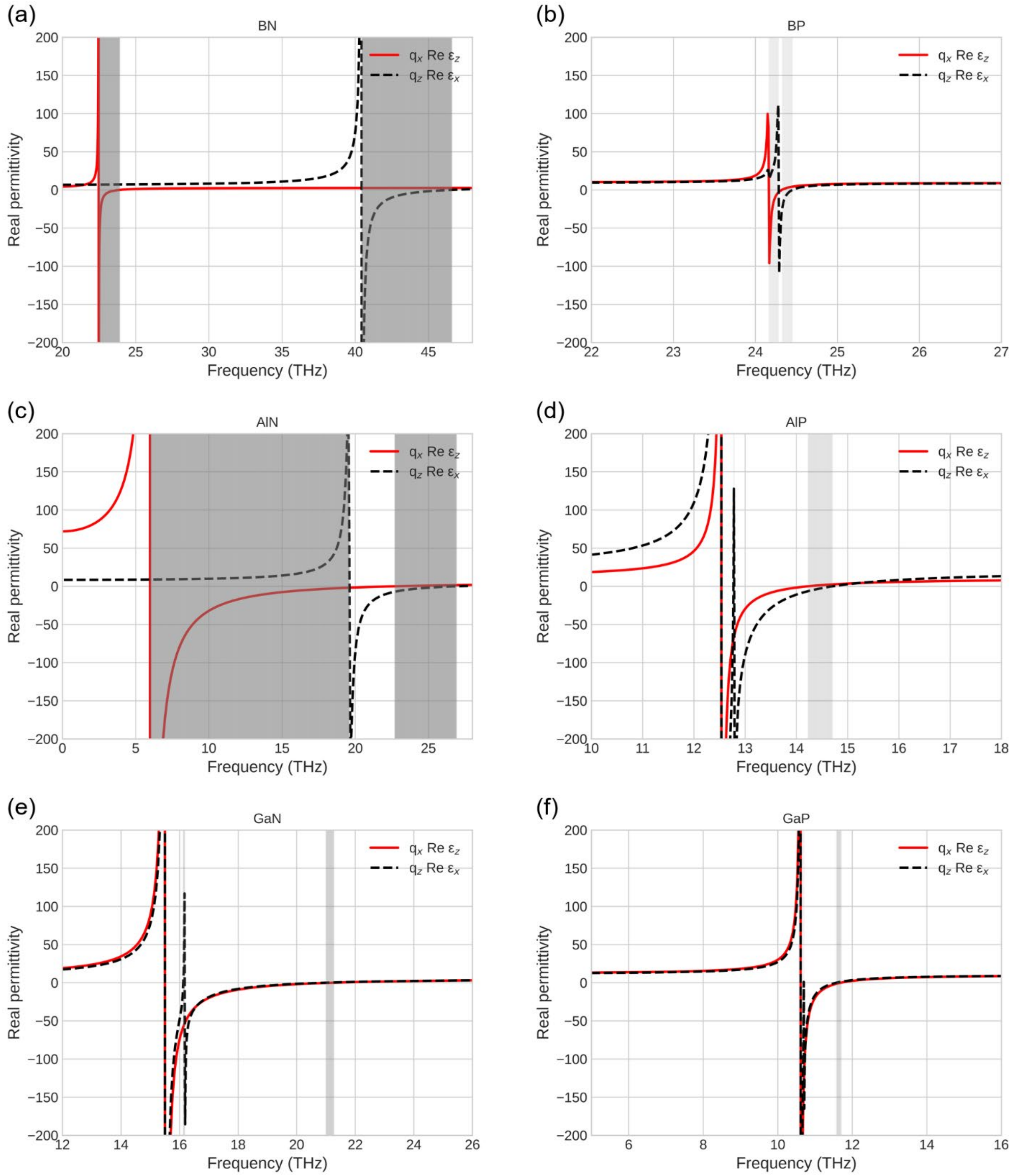

Fig. 6: Real part of dielectric functions for electromagnetic wave (phonon polariton) with wave vector in $x$ ($z$) direction and electric polarization in $z$ ($x$) direction in (a) h-BN, (b) h-BP, (c) h-AlN, (d) h-AlP, (e) h-GaN, and (f) h-GaP are represented with red solid (black dashed) lines. The grey regions are the hyperbolic bands.

Recent research has indicated that h-BN possesses hyperbolic PhP.[3,47–49] We hypothesize that similar materials, such as h-AlN, h-GaN, h-BP, h-AlP, and h-GaP, whose structures resemble h-BN, may also showcase hyperbolic PhP. To test our hypothesis, we computed the dielectric functions of these six materials.

As these materials are isotropic in the *x-y* plane, we only performed calculations in the *x* and *z* directions. In Equation (1), $\varepsilon_x$ ($\varepsilon_z$) is the dominant one when the wave vector is along the *z* (*x*) axis. Therefore, we plot $\varepsilon_x$ ($\varepsilon_z$) with the wave vector along the *z* (*x*) axis in Figure 6.

We found that h-BP, h-AlP, h-GaN, and h-GaP display relatively small frequency differences between the IR-active TO along the *x* and *z* directions in their phonon dispersion, as seen in Fig. 2b, 2d, 2e, and 2f. Thus, the peaks of the dielectric functions for different wave vector directions and polarization directions are close to each other, as depicted in Fig. 6b, 6d, 6e, and 6f. Consequently, the frequency regions where the dielectric function signs are opposite are minimal, almost eliminating the hyperbolic band windows.

In contrast, h-BN and h-AlN have a higher frequency difference between IR-active TO along the *x* and *z* directions in their phonon dispersion, as illustrated in Fig. 2a and 2c. Thus, the peaks of the dielectric functions for different wave vector directions and polarization directions are significantly separated, as shown in Fig. 6a and 6c. This leads to a larger frequency range where the dielectric function signs are opposite adjacent to the peaks, resulting in a more substantial hyperbolic band. Table 3 outlines the frequency ranges of the hyperbolic bands for h-BN and h-AlN, with h-AlN having a wider hyperbolic band range than h-BN.

Table.3: Hyperbolic bands of h-BN and h-AlN.

| | BN(typeI) | BN(typeII) | AlN(typeI) | AlN(typeII) |
|---|---|---|---|---|
| Hyperbolic band($cm^{-1}$) | 750-797 | 1349-1554 | 199-654 | 757-898 |

The reason for h-AlN having a wider hyperbolic band range than h-BN comes from Equation (20). One can see from Equation (20) that a smaller inverse of the phonon lifetime (γ, Figure 7) and phonon frequency ($\omega_{pk}$) results in a larger frequency region with a negative real part of the dielectric function at the right-hand side vicinity of that phonon frequency in the dielectric function real part curves (Figure 6), which leads to wider hyperbolic band range. The anisotropy of the crystal gives rise to the frequency difference of TO with different wave vector directions. The h-BN and h-AlN are stacked layers of planar structures with significant differences between the *x-y* directions and the *z* direction. On the other hand, the other four materials are wurtzite structures, with more minor differences between the *x-y* directions and the *z* direction. We can also observe this difference from their Born effective charge tensors (Table 4). The differences in the Born effective charge tensors of h-BN and h-AlN in the *x y* directions and the *z* direction are 1.77 and 0.63 electron charges, respectively, which are much larger than those of h-BP, h-AlP, h-GaN, and h-GaP, which are only 0.01, 0.15, 0.14, and 0.15 electron charges, respectively.

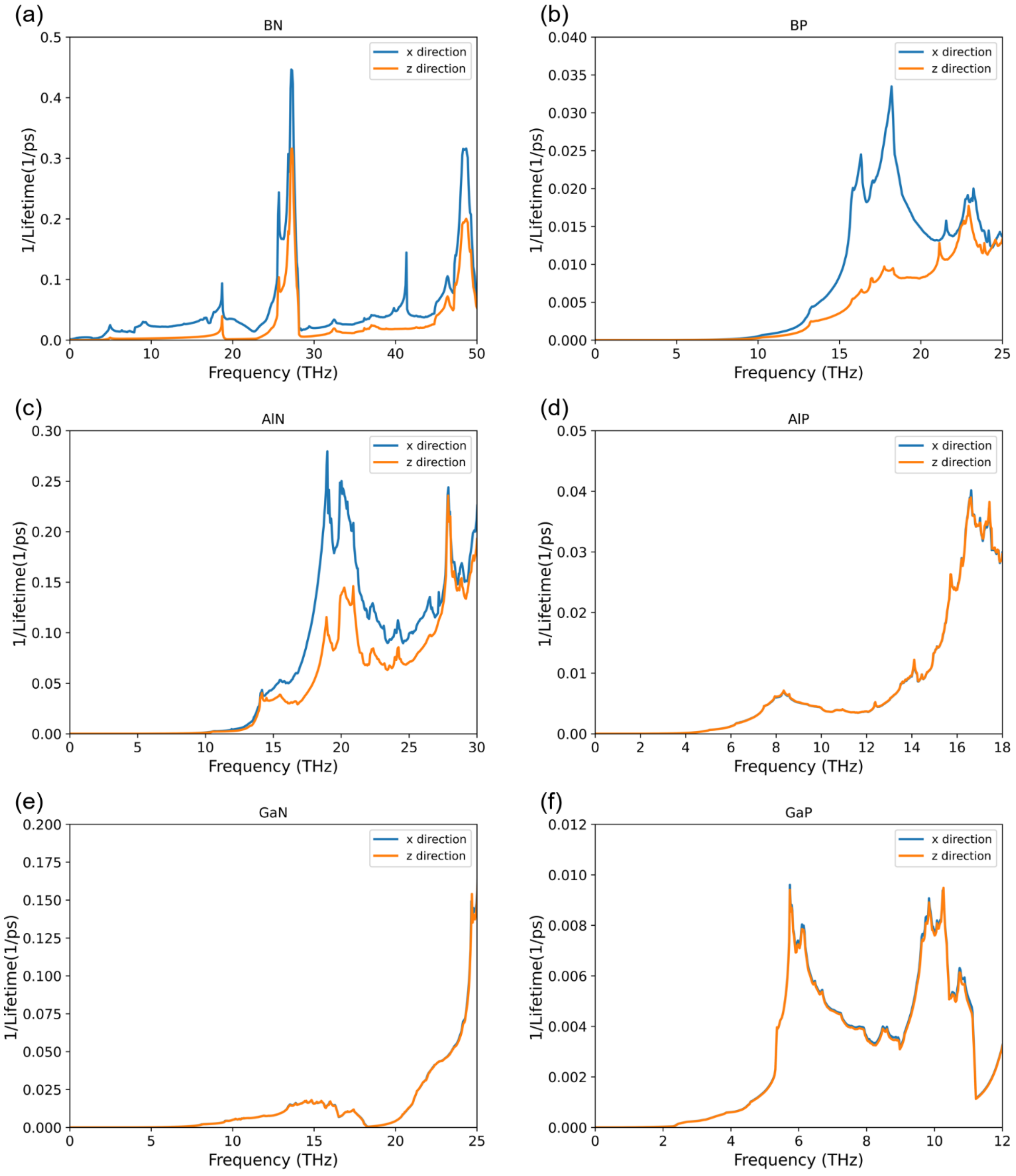

Fig. 7: The inverse of the phonon lifetime with wave vector in *x* and *z* direction for (a) h-BN, (b) h-BP, (c) h-AlN, (d) h-AlP, (e) h-GaN, and (f) h-GaP.

It is worth mentioning that when LO-TO splitting takes effect on the TO frequency, the dielectric function depends on the wave vector direction. Then, it becomes a rough estimation to determine whether a material is a hyperbolic material based on $\varepsilon_x$ and $\varepsilon_z$ with wave vector direction along x and z axis only. Instead, one should plot the isofrequency surface or curve (with Equation 30). Figure 8 shows examples. There is rarely a hyperbolic isofrequency curve in h-BP, h-AlP, and h-GaP (Figure 8b, 8d, and 8f), consistent with Figure 6b, 6d, and 6f, respectively. The frequency of the hyperbolic isofrequency curves of h-BN and h-AlN (Figure 8a and 8c) are inside or close to the hyperbolic bands in Figure 6a and 6c, respectively. These results indicate a good estimation of the hyperbolic band region by calculating the dielectric function with wave vector along the principal axis in most cases. On the other hand, there are some dumbbell-shaped and some butterfly-shaped isofrequency curves in h-AlN, h-GaP, and especially h-GaN (Figure 8c, 8f, and 8e). Compared to the dielectric function obtained without accounting for the wavevector direction at a fixed frequency, the dielectric function can undergo significant changes, and may even abruptly transit from positive to negative values, as the wavevector direction is varied. This gives rise to the characteristic dumbbell-shaped and butterfly-shaped isofrequency surfaces observed in certain materials. This means wave vector direction dependence of the dielectric function should be taken into account more seriously in these cases.

Table.4: Born effective charge tensor's nonzero elements (unit: electron charge)

| Material | Atom | $z_{xx}$ | $z_{yy}$ | $z_{zz}$ |
|---|---|---|---|---|
| BN | B | 2.72 | 2.72 | 0.95 |
| | N | -2.72 | -2.72 | -0.95 |
| BP | B | -0.57 | -0.57 | -0.56 |
| | P | 0.57 | 0.57 | 0.56 |
| AlN | Al | 2.52 | 2.52 | 3.15 |
| | N | -2.52 | -2.52 | -3.15 |
| AlP | Al | 2.20 | 2.20 | 2.35 |
| | P | -2.20 | -2.20 | -2.35 |
| GaN | Ga | 2.61 | 2.61 | 2.75 |
| | N | -2.61 | -2.61 | -2.75 |
| GaP | Ga | 2.05 | 2.05 | 2.20 |
| | P | -2.05 | -2.05 | -2.20 |

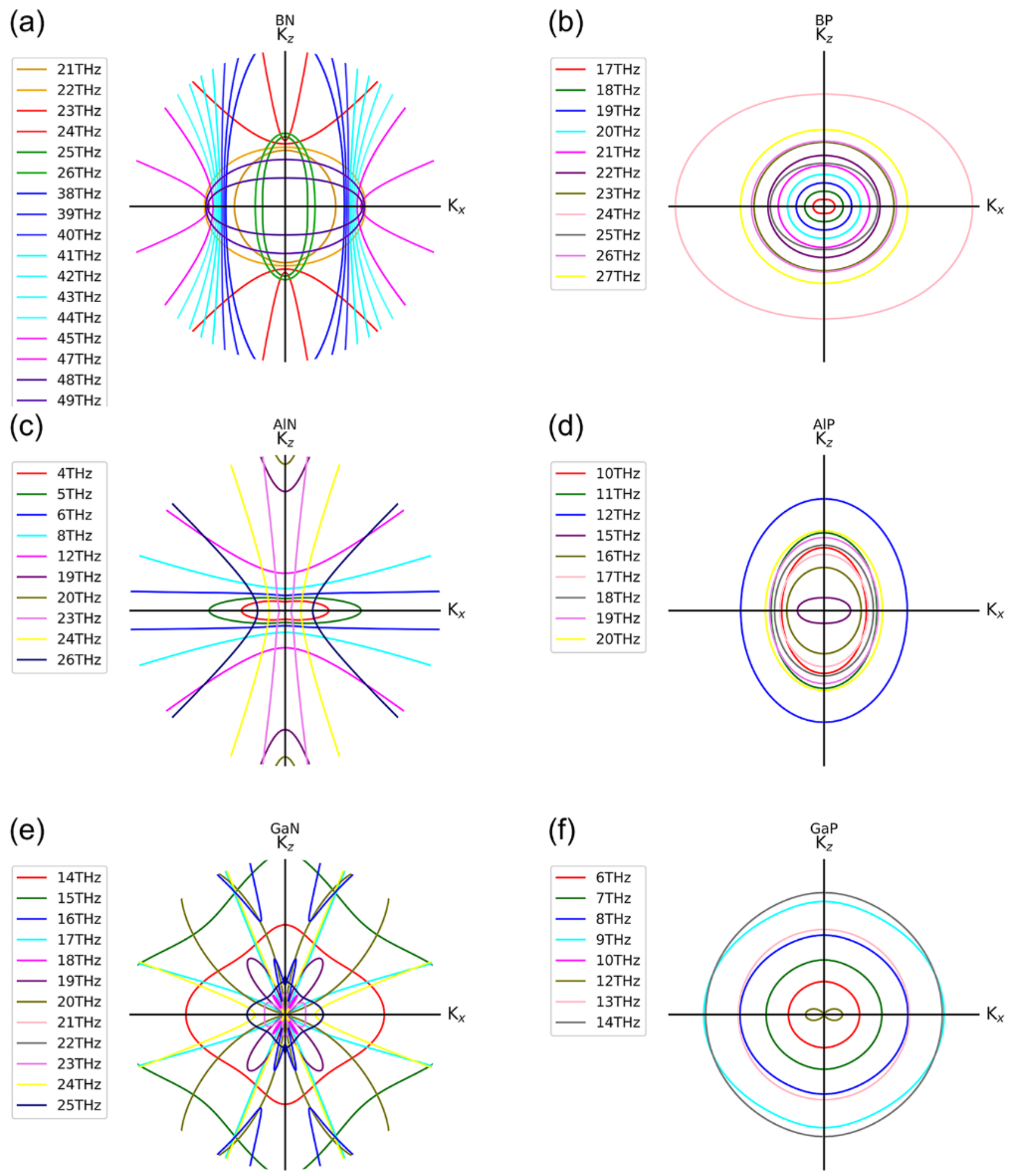

Fig. 8: Isofrequency curve of (a) h-BN, (b) h-BP, (c) h-AlN, (d) h-AlP, (e) h-GaN, and (f) h-GaP in x-z plane.

## 4. Conclusions

In this study, we considered the impact of LO-TO splitting on the dielectric function tensor of electromagnetic waves (PhPs) in different wave vector directions. We found that the dielectric function of h-BN obtained by considering LO-TO splitting is more consistent with experimental results than the usual approach where LO-TO splitting is not considered. Building on this, we investigated a series of hexagonal lattice III-V group semiconductors. We discovered that h-AlN also has a considerable range of hyperbolic bands, while the phosphides and GaN have almost no. The conditions for a material to have a wide hyperbolic band range are the significant difference in the IR-active TO frequencies in different wave vector directions, long phonon lifetimes, and low phonon frequencies. Dumbbell-shaped and butterfly-shaped isofrequency curves are found in h-AlN, h-GaP, and especially h-GaN.

## Notes

The authors declare no competing financial interests.

**Acknowledgements**

This work was supported by National Natural Science Foundation of China (no. 91963205); Science and Technology Planning Project of Guangdong Province (2023B1212060025) and Physical Research Platform (PRP) in School of Physics, SYSU.

## References

(1) Poddubny, A.; Iorsh, I.; Belov, P.; Kivshar, Y. Hyperbolic Metamaterials. *Nature Photonics* **2013**, *7* (12), 948–957. https://doi.org/10.1038/NPHOTON.2013.243.

(2) Hu, G.; Shen, J.; Qiu, C.; Alù, A.; Dai, S. Phonon Polaritons and Hyperbolic Response in van Der Waals Materials. *Adv. Optical Mater.* **2020**, *8* (5), 1901393. https://doi.org/10.1002/adom.201901393.

(3) Caldwell, J. D.; Kretinin, A. V.; Chen, Y.; Giannini, V.; Fogler, M. M.; Francescato, Y.; Ellis, C. T.; Tischler, J. G.; Woods, C. R.; Giles, A. J. et al. Novoselov, K. S. Sub-Diffractional Volume-Confined Polaritons in the Natural Hyperbolic Material Hexagonal Boron Nitride. *Nat Commun* **2014**, *5* (1), 5221. https://doi.org/10.1038/ncomms6221.

(4) Autore, M.; Li, P.; Dolado, I.; Alfaro-Mozaz, F. J.; Esteban, R.; Atxabal, A.; Casanova, F.; Hueso, L. E.; Alonso-González, P.; Aizpurua, J. et al. Hillenbrand, R. Boron Nitride Nanoresonators for Phonon-Enhanced Molecular Vibrational Spectroscopy at the Strong Coupling Limit. *Light Sci Appl* **2017**, *7* (4), 17172–17172. https://doi.org/10.1038/lsa.2017.172.

(5) Dunkelberger, A. D.; Ellis, C. T.; Ratchford, D. C.; Giles, A. J.; Kim, M.; Kim, C. S.; Spann, B. T.; Vurgaftman, I.; Tischler, J. G.; Long, J. P. et al. Active Tuning of Surface Phonon Polariton Resonances via Carrier Photoinjection. *NATURE PHOTONICS* **2018**, *12* (1), 50-+. https://doi.org/10.1038/s41566-017-0069-0.

(6) Salihoglu, H.; Xu, X. Near-Field Radiative Heat Transfer Enhancement Using Natural Hyperbolic Material. *Journal of Quantitative Spectroscopy and Radiative Transfer* **2019**, *222–223*, 115–121. https://doi.org/10.1016/j.jqsrt.2018.10.022.

(7) Zhang, R. Z. Optical and Thermal Radiative Properties of Topological Insulator Semiconductor Multilayers. *Journal of Quantitative Spectroscopy and Radiative Transfer* **2020**, *253*, 107133. https://doi.org/10.1016/j.jqsrt.2020.107133.

(8) Wu, X.; Fu, C.; Zhang, Z. Influence of hBN Orientation on the Near-Field Radiative Heat Transfer between Graphene/hBN Heterostructures. *Journal of Photonics for Energy* **2019**, *9* (3), 032702–032702.

(9) Lee, D.; Kim, M.; Lee, J.; Ko, B.; Park, H. J.; Rho, J. Angular Selection of Transmitted Light and Enhanced Spontaneous Emission in Grating-Coupled Hyperbolic Metamaterials. *Opt. Express* **2021**, *29* (14), 21458. https://doi.org/10.1364/OE.428231.

(10) Zhou, K.; Lu, L.; Li, B.; Cheng, Q. Hyperbolic Plasmon–Phonon Dispersion and Tunable Spontaneous Emission Enhancement in Ge2Sb2Te5-Based Multilayer Graphene and hBN System. *Journal of Applied Physics* **2021**, *130* (9), 093102. https://doi.org/10.1063/5.0058855.

(11) Guan, S.; Huang, S. Y.; Yao, Y.; Yang, S. A. Tunable Hyperbolic Dispersion and Negative Refraction in Natural Electride Materials. *Phys. Rev. B* **2017**, *95* (16), 165436. https://doi.org/10.1103/PhysRevB.95.165436.

(12) Sreekanth, K. V.; Simpson, R. E. Super-Collimation and Negative Refraction in Hyperbolic Van Der Waals Superlattices. *Optics Communications* **2019**, *440*, 150–154. https://doi.org/10.1016/j.optcom.2019.02.020.

(13) Cho, H.; Yang, Y.; Lee, D.; So, S.; Rho, J. Experimental Demonstration of Broadband Negative Refraction at Visible Frequencies by Critical Layer Thickness Analysis in a Vertical Hyperbolic Metamaterial. *Nanophotonics* **2021**, *10* (15), 3871–3877. https://doi.org/10.1515/nanoph-2021-0337.

(14) Wu, J.-S.; Basov, D. N.; Fogler, M. M. Topological Insulators Are Tunable Waveguides for Hyperbolic Polaritons. *Phys. Rev. B* **2015**, *92* (20), 205430. https://doi.org/10.1103/PhysRevB.92.205430.

(15) Feng, K.; Streyer, W.; Zhong, Y.; Hoffman, A. J.; Wasserman, D. Photonic Materials, Structures and Devices for Reststrahlen Optics. *OPTICS EXPRESS* **2015**, *23* (24), A1418–A1433. https://doi.org/10.1364/OE.23.0A1418.

(16) Lee, D.; So, S.; Hu, G.; Kim, M.; Badloe, T.; Cho, H.; Kim, J.; Kim, H.; Qiu, C.-W.; Rho, J. Hyperbolic Metamaterials: Fusing Artificial Structures to Natural 2D Materials. *eLight* **2022**, *2* (1), 1. https://doi.org/10.1186/s43593-021-00008-6.

(17) Lee, M.; Lee, E.; So, S.; Byun, S.; Son, J.; Ge, B.; Lee, H.; Park, H. S.; Shim, W.; Pee, J. H. et al. Bulk Metamaterials Exhibiting Chemically Tunable Hyperbolic Responses. *J. Am. Chem. Soc.* **2021**, *143* (49), 20725–20734. https://doi.org/10.1021/jacs.1c08446.

(18) Lee, M.; Lee, E.; Byun, S.; Kim, J.; Yun, J.; So, S.; Lee, H.; Pee, J. H.; Shim, W.; Cho, S.-P. et al. R-BN: A Fine Hyperbolic Dispersion Modulator for Bulk Metamaterials Consisting of Heterostructured Nanohybrids of h-BN and Graphene. *Journal of Solid State Chemistry* **2022**, *309*, 122937. https://doi.org/10.1016/j.jssc.2022.122937.

(19) Ambrosio, A.; Jauregui, L. A.; Dai, S.; Chaudhary, K.; Tamagnone, M.; Fogler, M. M.; Basov, D. N.; Capasso, F.; Kim, P.; Wilson, W. L. Mechanical Detection and Imaging of Hyperbolic Phonon Polaritons in Hexagonal Boron Nitride. *ACS nano* **2017**, *11* (9), 8741–8746.

(20) Ma, W.; Alonso-Gonzalez, P.; Li, S.; Nikitin, A. Y.; Yuan, J.; Martin-Sanchez, J.; Taboada-Gutierrez, J.; Amenabar, I.; Li, P.; Velez, S. et al. In-Plane Anisotropic and Ultra-Low-Loss Polaritons in a Natural van Der Waals Crystal. *NATURE* **2018**, *562* (7728), 557-+. https://doi.org/10.1038/s41586-018-0618-9.

(21) Ma, W.; Hu, G.; Hu, D.; Chen, R.; Sun, T.; Zhang, X.; Dai, Q.; Zeng, Y.; Alù, A.; Qiu, C.-W. et al. Ghost Hyperbolic Surface Polaritons in Bulk Anisotropic Crystals. *Nature* **2021**, *596* (7872), 362–366. https://doi.org/10.1038/s41586-021-03755-1.

(22) Álvarez-Pérez, G.; Folland, T. G.; Errea, I.; Taboada-Gutiérrez, J.; Duan, J.; Martín-Sánchez, J.; Tresguerres-Mata, A. I. F.; Matson, J. R.; Bylinkin, A.; He, M. et al. Infrared Permittivity of the Biaxial van Der Waals Semiconductor α-$MoO_3$ from Near- and Far-Field Correlative Studies. *Advanced Materials* **2020**, *32* (29), 1908176. https://doi.org/10.1002/adma.201908176.

(23) Taboada-Gutiérrez, J.; Álvarez-Pérez, G.; Duan, J.; Ma, W.; Crowley, K.; Prieto, I.; Bylinkin, A.; Autore, M.; Volkova, H.; Kimura, K. et al. Broad Spectral Tuning of Ultra-Low-Loss Polaritons in a van Der Waals Crystal by Intercalation. *Nat. Mater.* **2020**, *19* (9), 964–968. https://doi.org/10.1038/s41563-020-0665-0.

(24) Tong, Z.; Dumitrică, T.; Frauenheim, T. First-Principles Prediction of Infrared Phonon and Dielectric Function in Biaxial Hyperbolic van Der Waals Crystal α-$MoO_3$. *Phys. Chem. Chem. Phys.* **2021**, *23* (35), 19627–19635. https://doi.org/10.1039/D1CP00682G.

(25) Ramesh, M.; Niranjan, M. K. Phonon Modes, Dielectric Properties, Infrared Reflectivity, and Raman Intensity Spectra of Semiconducting Silicide BaSi2: First Principles Study. *Journal of Physics and Chemistry of Solids* **2018**, *121*, 219–227. https://doi.org/10.1016/j.jpcs.2018.05.033.

(26) Gonze, X.; Lee, C. Dynamical Matrices, Born Effective Charges, Dielectric Permittivity Tensors, and Interatomic Force Constants from Density-Functional Perturbation Theory. *Phys. Rev. B, Condens. Matter (USA)* **1997**, *55* (16), 10355–10368. https://doi.org/10.1103/PhysRevB.55.10355.

(27) Rivera, N.; Coulter, J.; Christensen, T.; Narang, P. Ab Initio Calculation of Phonon Polaritons in Silicon Carbide and Boron Nitride. *arXiv:1809.00058 [cond-mat, physics:physics, physics:quant-ph]Submitted on* 31 Aug **2018**. https://arxiv.org/abs/1809.00058

(28) De Oliveira, T. V. A. G.; Nörenberg, T.; Álvarez-Pérez, G.; Wehmeier, L.; Taboada-Gutiérrez, J.; Obst, M.; Hempel, F.; Lee, E. J. H.; Klopf, J. M.; Errea, I. et al. Nanoscale-Confined Terahertz Polaritons in a van Der Waals Crystal. *Advanced Materials* **2021**, *33* (2), 2005777. https://doi.org/10.1002/adma.202005777.

(29) Di Sia, P. Overview of Drude-Lorentz Type Models and Their Applications. *Nanoscale Systems: Mathematical Modeling, Theory and Applications* **2014**, *3*, 1–13. https://doi.org/10.2478/nsmmt-2014-0001.

(30) Duan, J.; Chen, R.; Li, J.; Jin, K.; Sun, Z.; Chen, J. Launching Phonon Polaritons by Natural Boron Nitride Wrinkles with Modifiable Dispersion by Dielectric Environments. *Advanced Materials* **2017**, *29* (38), 1702494. https://doi.org/10.1002/adma.201702494.

(31) Breslin, V. M.; Ratchford, D. C.; Giles, A. J.; Dunkelberger, A. D.; Owrutsky, J. C. Hyperbolic Phonon Polariton Resonances in Calcite Nanopillars. *Opt. Express* **2021**, *29* (8), 11760. https://doi.org/10.1364/OE.417405.

(32) Dai, S.; Fang, W.; Rivera, N.; Stehle, Y.; Jiang, B.-Y.; Shen, J.; Tay, R. Y.; Ciccarino, C. J.; Ma, Q.; Rodan-Legrain, D. Phonon Polaritons in Monolayers of Hexagonal Boron Nitride. *Advanced materials* **2019**, *31* (37), 1806603.

(33) Ma, W.; Shabbir, B.; Ou, Q.; Dong, Y.; Chen, H.; Li, P.; Zhang, X.; Lu, Y.; Bao, Q. Anisotropic Polaritons in van Der Waals Materials. *InfoMat* **2020**, *2* (5), 777–790.

(34) Dai, S.; Fei, Z.; Ma, Q.; Rodin, A.; Wagner, M.; McLeod, A.; Liu, M.; Gannett, W.; Regan, W.; Watanabe, K. et al. Tunable Phonon Polaritons in Atomically Thin van Der Waals Crystals of Boron Nitride. *SCIENCE* **2014**, *343* (6175), 1125–1129. https://doi.org/10.1126/science.1246833.

(35) Huber, A.; Ocelic, N.; Kazantsev, D.; Hillenbrand, R. Near-Field Imaging of Mid-Infrared Surface Phonon Polariton Propagation. *Applied Physics Letters* **2005**, *87* (8), 081103. https://doi.org/10.1063/1.2032595.

(36) Gonze, X.; Charlier, J.-C.; Allan, D. C.; Teter, M. P. Interatomic Force Constants from First Principles: The Case of \ensuremath{\alpha}-Quartz. *Phys. Rev. B* **1994**, *50* (17), 13035–13038. https://doi.org/10.1103/PhysRevB.50.13035.

(37) Giannozzi, P.; de Gironcoli, S.; Pavone, P.; Baroni, S. Ab Initio Calculation of Phonon Dispersions in Semiconductors. *Phys. Rev. B* **1991**, *43* (9), 7231–7242. https://doi.org/10.1103/PhysRevB.43.7231.

(38) Kresse; Furthmuller. Efficient Iterative Schemes for Ab Initio Total-Energy Calculations Using a Plane-Wave Basis Set. *Physical review. B, Condensed matter* **1996**, *54* (16), 11169–11186. https://doi.org/10.1103/PhysRevB.54.11169.

(39) Kresse, G.; Joubert, D. From Ultrasoft Pseudopotentials to the Projector Augmented-Wave Method. *PHYSICAL REVIEW B* **1999**, *59* (3), 1758–1775. https://doi.org/10.1103/PhysRevB.59.1758.

(40) Kresse, G.; Furthmüller, J. Efficiency of Ab-Initio Total Energy Calculations for Metals and Semiconductors Using a Plane-Wave Basis Set. *Computational Materials Science* **1996**, *6* (1), 15–50. https://doi.org/10.1016/0927-0256(96)00008-0.

(41) Togo, A. First-Principles Phonon Calculations with Phonopy and Phono3py. *J. Phys. Soc. Jpn.* **2023**, *92* (1), 012001. https://doi.org/10.7566/JPSJ.92.012001.

(42) Togo, A.; Chaput, L.; Tadano, T.; Tanaka, I. Implementation Strategies in Phonopy and Phono3py. *J. Phys.: Condens. Matter* **2023**, *35* (35), 353001. https://doi.org/10.1088/1361-648X/acd831.

(43) Perdew; Burke; Ernzerhof. Generalized Gradient Approximation Made Simple. *Physical review letters* **1996**, *77* (18), 3865–3868. https://doi.org/10.1103/PhysRevLett.77.3865.

(44) Michel, K. H.; Verberck, B. Phonon Dispersions and Piezoelectricity in Bulk and Multilayers of Hexagonal Boron Nitride. *Phys. Rev. B* **2011**, *83* (11), 115328. https://doi.org/10.1103/PhysRevB.83.115328.

(45) Davydov, V. Yu.; Kitaev, Yu. E.; Goncharuk, I. N.; Smirnov, A. N.; Graul, J.; Semchinova, O.; Uffmann, D.; Smirnov, M. B.; Mirgorodsky, A. P.; Evarestov, R. A. Phonon Dispersion and Raman Scattering in Hexagonal GaN and AlN. *Phys. Rev. B* **1998**, *58* (19), 12899–12907. https://doi.org/10.1103/PhysRevB.58.12899.

(46) Deng, Z.; Li, Z.; Wang, W. Electron Affinity and Ionization Potential of Two-Dimensional Honeycomb Sheets: A First Principle Study. *Chemical Physics Letters* **2015**, *637*, 26–31. https://doi.org/10.1016/j.cplett.2015.07.054.

(47) Zhou, S.; Khan, A.; Fu, S.-F.; Wang, X.-Z. Extraordinary Reflection and Refraction from Natural Hyperbolic Materials. *Opt. Express* **2019**, *27* (11), 15222. https://doi.org/10.1364/OE.27.015222.

(48) Viner, J. J. S.; McDonnell, L. P.; Rivera, P.; Xu, X.; Smith, D. C. Insights into Hyperbolic Phonon Polaritons in h − BN Using Raman Scattering from Encapsulated Transition Metal Dichalcogenide Layers. *Phys. Rev. B* **2021**, *104* (16), 165404. https://doi.org/10.1103/PhysRevB.104.165404.

(49) Dai, S.; Ma, Q.; Andersen, T.; Mcleod, A. S.; Fei, Z.; Liu, M. K.; Wagner, M.; Watanabe, K.; Taniguchi, T.; Thiemens, M.; Keilmann, F. et al. Subdiffractional Focusing and Guiding of Polaritonic Rays in a Natural Hyperbolic Material. *Nat Commun* **2015**, *6* (1), 6963. https://doi.org/10.1038/ncomms7963.